\documentstyle[12pt]{article}
\normalbaselines

\begin{document}
\begin{flushright}

{\raggedleft hep-th/9311099\\[2.cm]}

\end{flushright}

\renewcommand{\thefootnote}{\fnsymbol{footnote}}
\begin{center}
{\LARGE\baselineskip1.1cm Nonlocal Quantum Electrodynamics
Admits a Finitely Induced Gauge Field Action\\[2.cm]}
{\large K. Scharnhorst\footnote[1]{E-mail:
kjsch @ qft.physik.uni-leipzig.d400.de}}\\[0.3cm]
{\small Berlin,
Federal Republic of Germany}\\[3.cm]
\thispagestyle{empty}
\end{center}
\renewcommand{\thefootnote}{\arabic{footnote}}

\begin {abstract}
\noindent
The Letter reconsiders a result obtained by {\sc Chr\'etien}
and {\sc Peierls} in 1954 within nonlocal QED in 4D [Proc.
Roy. Soc. London {\bf A 223}, 468]. Starting from secondly
quantized fermions subject to a nonlocal action with the kernel
$\left[ i\not\hspace{-0.07cm}\partial_x\ a\left(x\right)
-\ m\ b\left(x\right)\right]$ and gauge covariantly coupled
to an external U(1) gauge field they found that for $a = b$ the
induced gauge field action cannot be made finite irrespectively
of the choice of the nonlocality $a$ $(= b)$.
But, the general case $a \neq b$
naturally to be studied admits a finitely induced gauge field
action, as the present Letter demonstrates.\\[0.6cm]
\end{abstract}

\newpage
\noindent
Despite its inherent difficulties nonlocal quantum
field theory has received continued although
changing attention in past decades (for a review and
references see \cite{efim1}--\cite{efim3}).
Restricting ourselves to gauge theories
let us mention that in recent years work in
nonlocal theories has been performed within
quantum field theory \cite{part}--\cite{evan}
as well as in the neighbouring field of the study of
effective Lagrangians in nuclear theory
\cite{ohta}--\cite{tern}. The conceptual intentions
in the few references selected are quite diverse, but
they have in common to refer and to relate to earlier work of
{\sc Chr\'etien} and {\sc Peierls} dealing with nonlocal
QED in 4D \cite{f}. This investigation explored a
special Ansatz for nonlocal QED with the aim to
remove the UV divergencies present in standard, local
QED (for the Ansatz see Eq.\ (\ref{D3}) below,
the special choice of {\sc Chr\'etien} and {\sc Peierls}
was $a = b$, $a (=b)$ arbitrary else).
But, the attempt failed (see also \cite{peie}).
In particular, this negative result is believed to have
closed the door to a finitely induced gauge field action.
However, because {\sc Chr\'etien} and {\sc Peierls}
extended their consideration not to
the most general setting possible ($a \neq b$),
this belief is not correct in general.
Contrary to this belief, in this Letter we will demonstrate that
nonlocal QED in 4D admits a finitely induced gauge field
action under appropriate choice of the nonlocal Ansatz applied.
The purpose of the present Letter therefore
is a purely technical one. Conceptual questions the
calculation below grew from are dealt with in
a separate comprehensive paper \cite{scha}.\\

Consider within QED in 4D Minkowski space
the following gauge covariant fermion action.
\parindent0.em

\begin{eqnarray}
\label{D3}
\Gamma_{F} [A,\psi,\bar\psi]\ &=&\ \int d^4x\ d^4x^\prime\ \ \bar\psi (x)
\ \ {\rm e}^{\displaystyle\ ie \int^{x^\prime}_x dy_\mu\ A^\mu(y)}\ \
\cdot\nonumber\\
\vspace{0.3cm}\nonumber\\
&&\hspace{-1.cm}\cdot\ \left[ a\left(x-x^\prime\right)
\ \left( i\not\hspace{-0.07cm}\partial_{x^\prime}
- e \not\hspace{-0.13cm}A(x^\prime) \right)\
-\ m\ b\left(x-x^\prime\right) \right]\psi (x^\prime)
\end{eqnarray}

$a$, $b$ are functions (distributions) arbitrary for the moment.
The line integration in the phase factor
is understood to be performed along a
straight line connecting starting and end point.
Eq.\ (\ref{D3}) is written in such shape as to keep
contact with standard (local) QED ($\tilde a\ =\ \tilde b\ \equiv\ 1$)
\footnote{Fourier transforms are defined for $f(x)$
by ($f$ stands for $a$, $b$.)

\begin{eqnarray*}
\tilde f(p)\ =\ \int d^4x\ \ {\rm e}^{\ -ipx}\ \ f(x)\ \ \ .
\end{eqnarray*}
For convenience, we always write
$\tilde f(s)$ instead of $\tilde f(p)$ for one
and the same function $\tilde f$
because all functions $f(x)$ we are studying depend
on $x$ via $x^2$ only ($s = - p^2/ m^2 = p^2_E/ m^2$; the subscript $E$
refers to the (Wick rotated) Euclidean momentum variable).}
as close as possible (When referring below to standard QED
we always have in mind local QED).\\

\parindent1.5em

The object of the study is the induced gauge field
action $\Gamma_G [A]$ given by the formula

\parindent0.em
\begin{eqnarray}
\label{ZA1}
{\rm e}^{\displaystyle\ i \Gamma_G [A]}\ \ &=&\
\int D\psi D\bar\psi\
\ {\rm e}^{\displaystyle\ i\Gamma_F [A,\psi,\bar\psi]}\ \ \ .
\end{eqnarray}

We here report on the calculation of the
coefficients of the first two quadratic terms of the
derivative expansion of $\Gamma_G [A]$, i.e.,
the coefficient of the mass term $A_\mu A^\mu$ and the
coefficients of $(\partial_\mu A^\mu)^2$ and
$\partial_\mu A_\nu \partial^\mu A^\nu$, which are
divergent in standard QED.
All further terms which from standard QED are known
to be finite even in the local limit
$\tilde a\ =\ \tilde b\ \equiv\ 1$ are considered beyond
present interest and will be commented at the end of the Letter only.
We will deliberately employ a gauge non-invariant regularization
in order to also study the behaviour of the mass term coefficient when
lifting the regularization.\\

\parindent1.5em

For the purpose of the explicit calculation we rewrite
$\Gamma_F [A,\psi,\bar\psi]$ in
the following symmetrized form.
\parindent0.em

\begin{eqnarray}
\label{ZA2}
&&\hspace{-1.cm}\Gamma_F [A,\psi,\bar\psi]\ =\nonumber\\
\vspace{0.3cm}\nonumber\\
&&\hspace{-1.cm}=\ {1\over2}\ \int d^4x\ d^4x^\prime\ \ \bar\psi (x)
\ \ {\rm e}^{\displaystyle\ ie \int^{x^\prime}_x dy_\mu\ A^\mu(y)}\ \
\cdot\nonumber\\
\vspace{0.3cm}\nonumber\\
&&\hspace{-0.2cm}\cdot\ \left[ a\left( x-x^\prime\right)
\ \left( i\stackrel{\rightarrow}{\not\hspace{-0.07cm}\partial_{x^\prime}}
- e \not\hspace{-0.13cm}A(x^\prime) \right)\
-\ m\ b\left( x-x^\prime\right) \right]\psi (x^\prime)\ + \nonumber\\
\vspace{0.3cm}\nonumber\\
&&\hspace{-0.3cm} +\ {1\over2}\ \int d^4x\ d^4x^\prime\ \ \bar\psi (x)
\ \left[ -\left( i
\stackrel{\leftarrow}{\not\hspace{-0.07cm}\partial_x}
+ e \not\hspace{-0.13cm}A(x) \right)\
\ a\left( x-x^\prime\right)\
-\ m\ b\left( x-x^\prime\right) \right]\ \
\cdot\nonumber\\
\vspace{0.3cm}\nonumber\\
&&\hspace{0.8cm}\cdot
\ {\rm e}^{\displaystyle\ ie \int^{x^\prime}_x dy_\mu\ A^\mu(y)}\
\psi (x^\prime)
\end{eqnarray}

We then expand the right hand side of Eq.\ (\ref{ZA2}) in
powers of $A_\mu$ up to $O(A^2)$ (i.e., $O(e^2)$) and
insert following expansions (the upper obtained by using
$y_\mu(\tau)\ =\ (x^\prime - x)_\mu\ \tau + x_\mu$,
$\tau\ \in\ [0,1]$).

\begin{eqnarray}
\label{ZA3}
&&\hspace{-1.5cm}\int^{x^\prime}_x dy_\mu\ A^\mu(y)\ =\
(x^\prime - x)^\mu\ \bigg\{\ A_\mu(y)\ +\nonumber\\
&&\ +\ \left. {1\over 24}\ (x^\prime - x)^\nu
(x^\prime - x)^\lambda
\ \partial_\nu \partial_\lambda\ A_\mu(y)
\ +\ \ldots\ \right\}_{y={(x+x^\prime)\over 2}}\ \ \ ,\\
\vspace{0.3cm}\nonumber\\
\label{ZA4}
&&\hspace{-1.5cm} A_\mu (x)\ +\ A_\mu (x^\prime)\ =\ \nonumber\\
&&=\ 2\ \left\{\ A_\mu (y)\ +\
{1\over 8}\ (x^\prime - x)^\nu(x^\prime - x)^\lambda\
\partial_\nu \partial_\lambda\ A_\mu (y)
\ +\ \ldots\
\right\}_{y={(x+x^\prime)\over 2}}\  .
\end{eqnarray}

For calculating the coefficients of $A_\mu A^\mu$,
$(\partial_\mu A^\mu)^2$, and
$\partial_\mu A_\nu \partial^\mu A^\nu$ in $\Gamma_G [A]$
it is sufficient to keep at most
two derivatives acting on the gauge potentials
in $\Gamma_F [A,\psi,\bar\psi]$.
The expression obtained this way for
$\Gamma_F$ (we will not give
this rather long expression) now serves as the
starting point for deriving Feynman rules and
calculating the effective action terms desired.
One should take notice that $\Gamma_F$ also contains
terms quadratic in $A_\mu$ what leads to the
situation that besides the standard photon polarization
diagram also a tadpole contribution to the
photon self-energy is to be taken into account.\\

\parindent1.5em

The explicit calculation of the terms we are aiming
at is quite tedious and shall not be displayed here.
We only comment few points of the calculation.
Coordinate differences as occurring in Eqs.\ (\ref{ZA3}),
(\ref{ZA4}) are translated into momentum space as
derivatives with respect to a corresponding momentum variable
acting on certain functions in momentum space. This of
course involves partial integrations in momentum space
for which as usual boundary contributions are assumed not to occur.
The photon polarization function is a nonlocal distribution. Therefore,
from the formal expression derived by the Feynman rules
the local structures we are interested in have to be
extracted. In order to properly define this procedure
we apply a (radial) momentum space UV cut-off at $\Lambda$
for the loop integration. This regularization is
most suited for our purposes. The final result will be given within
this gauge non-invariant cut-off regularization.
Furthermore, a Wick rotation for the loop integration
is performed and such equivalences like (\ref{DF5}),
(\ref{DF6}) further below are used. Then, the final result reads
\parindent0.em

\begin{eqnarray}
\label{ZA5}
&&\hspace{-0.5cm}\Gamma_G [A]\ =\ const.\ +\
{e^2\over 16\pi^2}
\ \int d^4x\ \bigg\{\ C_0\ m^2 A_\mu(x) A^\mu(x)\ + \nonumber\\
\vspace{0.3cm}\nonumber\\
&&+\ \Big[\ C_{1s}\
[g_{\mu\nu} g_{\alpha\beta}\ +\ g_{\mu\alpha} g_{\nu\beta}]\ +\
C_{1a}\ [g_{\mu\nu} g_{\alpha\beta}\ -\ g_{\mu\alpha} g_{\nu\beta}]\
\Big]\ A^\mu(x) \partial^\alpha \partial^\beta A^\nu(x)\ +
\nonumber\\
\vspace{0.3cm}\nonumber\\
&&+\ \ldots \bigg\}
\end{eqnarray}

where ($f^\prime= d/ds\ f$)

\begin{eqnarray}
\label{ZA6}
C_0\ \ &=&\ -\ s^2\ h^\prime\ \ \Bigg\vert_0^{\Lambda^2\over m^2}\ \ \ \ ,\\
\vspace{1.2cm}\nonumber\\
\label{ZA7}
\hspace{-1.cm}C_{1s}\ &=&\ -\ {1\over 6}\ s^3\ h^{\prime\prime\prime}\ -\
{1\over 2}\ s^2\ h^{\prime\prime}\ +\nonumber\\
\vspace{0.2cm}\nonumber\\
&&\ +\ {1\over 2}\ \left(\ {\rm e}^{\displaystyle -h}\ \left[\
s^4\ \tilde a \tilde a^{\prime\prime}\ +\
2\ s^3\ \tilde a \tilde a^\prime\ +\
s^3\ \tilde b \tilde b^{\prime\prime}\ +\
s^2\ \tilde b \tilde b^\prime\ \right]\right)^\prime\ -\nonumber\\
\vspace{0.2cm}\nonumber\\
&&\ -\ {\rm e}^{\displaystyle -h}\
\left[\ {1\over 3}\ s^4\ \tilde a \tilde a^{\prime\prime\prime}\ +\
2\ s^3\ \tilde a \tilde a^{\prime\prime}\ +\
{1\over 3}\ s^3\ \tilde b \tilde b^{\prime\prime\prime}\ +\
2\ s^2\ \tilde a \tilde a^\prime\ +\right.\nonumber\\
&&\hspace{1.8cm}
+\left. {3\over 2}\ s^2\ \tilde b \tilde b^{\prime\prime}\ +\
s\ \tilde b \tilde b^\prime\ \right]\ \ \ \Bigg\vert_0^{\Lambda^2\over m^2}
\ \ \ \ ,\\
\vspace{1.2cm}\nonumber\\
\label{ZA8}
C_{1a}\ &=&\ {1\over 18}\ s^3\ h^{\prime\prime\prime}\ -\
{1\over 6}\ s^2\ h^{\prime\prime}\ -\
{2\over 3}\ s\ h^\prime\ +\
{2\over 3}\ h\ +\nonumber\\
\vspace{0.2cm}\nonumber\\
&&\ +\ {1\over 2}\ \left(\ {\rm e}^{\displaystyle -h}\ \left[\
-\ {1\over 3}\ s^4\ \tilde a \tilde a^{\prime\prime}\ +\
{2\over 3}\ s^3\ \tilde a \tilde a^\prime\ -\
{1\over 3}\ s^3\ \tilde b \tilde b^{\prime\prime}\ +\
s^2\ \tilde b \tilde b^\prime\ \right]\right)^\prime\ +\nonumber\\
\vspace{0.2cm}\nonumber\\
&&\ +\ {\rm e}^{\displaystyle -h}\
\left[\ {1\over 9}\ s^4\ \tilde a \tilde a^{\prime\prime\prime}\ +\
{4\over 3}\ s^3\ \tilde a \tilde a^{\prime\prime}\ +\
{1\over 9}\ s^3\ \tilde b \tilde b^{\prime\prime\prime}\ +\
2\ s^2\ \tilde a \tilde a^\prime\ +\right.\nonumber\\
&&\hspace{1.8cm}+\left.\
{7\over 6}\ s^2\ \tilde b \tilde b^{\prime\prime}\ +\
2\ s\ \tilde b \tilde b^\prime\ \right]\ \ \
\Bigg\vert_0^{\Lambda^2\over m^2}\ - \nonumber\\
&&\ -\ \int\limits_0^{\Lambda^2\over m^2} ds\
{1\over s \tilde a^2 + \tilde b^2}\
\left[\ {s\ \tilde a^2 \over s \tilde a^2 + \tilde b^2}\
\left[\ s\ \tilde a \tilde a^\prime + \tilde b \tilde b^\prime\ \right]
\ +\right.\nonumber\\
\vspace{0.2cm}\nonumber\\
&&\hspace{1.5cm}
+\ {2\over 3}\ s^3\ \tilde a \tilde a^{\prime\prime\prime}\ +\
3\ s^2\ \tilde a \tilde a^{\prime\prime}\ +\
{2\over 3}\ s^2\ \tilde b \tilde b^{\prime\prime\prime}\ +\nonumber\\
&&\hspace{1.5cm}+\ 2\ s\ \tilde a \tilde a^\prime\ +\
3\ s\ \tilde b \tilde b^{\prime\prime}\ -\
s\ (\tilde b^\prime)^2\ +\
3\ \tilde b \tilde b^\prime\ \Bigg] \ \ \ ,\\
\vspace{0.3cm}\nonumber\\
h\ &=&\ h(s)\ =\ \ln\left[ s \tilde a^2 +\tilde b^2 \right]
\hspace{0.5cm},\hspace{0.5cm}\tilde a\ = \tilde a (s)
\hspace{0.5cm},\hspace{0.5cm}\tilde b\ = \tilde b (s)\ \ \ .\nonumber
\end{eqnarray}

The displayed result is exact for any value of the cut-off $\Lambda$,
so far no term vanishing at removing the cut-off has been
neglected. For $\tilde a\ =\ \tilde b\ \equiv\ 1$ the
standard QED result is reproduced (cf. \cite{ach}; \cite{jau},
Eq.\ (9-64), for $\Lambda \longrightarrow \infty$ the coefficient $C(0)$
there is related to our expressions by the equation
$C(0) = - e^2\ (5 C_{1s} + 3 C_{1a})/24\pi^2$).\\
\parindent1.5em

The expression for the mass term coefficient (\ref{ZA6})
can easily be rederived by an independent method.
In order to look for a mass term of the gauge field $A_\mu$
we can restrict ourselves to the class of constant gauge potentials
$A_\mu (x)\ =\ e^{-1}k_\mu\ \equiv\ const.$ the consideration of which
is sufficient for this purpose. For this simple background
$\Gamma_G  [A]$ is given by the determinant of the kernel of
the fermion action $\Gamma_F$ in the presence of the constant background
$k_\mu$ which can be viewed in momentum space representation
as a constant external momentum.
The induced gauge field action reads then
$\left( h\ =\ \ln\left[ s \tilde a^2 +\tilde b^2 \right]\right)$
\parindent0.em

\parindent0.em
\begin{eqnarray}
\label{DF1}
\Gamma_G [e^{-1} k]\ &=&\ -\ 2i\ V_4 \int\limits_\Lambda
{d^4p\over (2\pi)^4}\ \
h\left(-(p+k)^2\over m^2 \right)\ \ \ .
\end{eqnarray}

The subscript $\Lambda$ in Eq.\ (\ref{DF1}) again
indicates that we apply a cut-off regularization with a (radial)
momentum space UV cut-off at $\Lambda$. Because we cannot assume
from the very beginning that the result in Eq.\ (\ref{DF1})
will be finite (this is related to the vacuum energy
problem which we will not consider) we
are barred from simply using a shift $p\longrightarrow p-k$ (which
would make vanish the dependence on $k$ at once; this would only
be applicable in a gauge invariant regularization).\\
\parindent1.5em

Let us further transform the integral appearing
in Eq.\ (\ref{DF1}). First, we perform a Wick rotation
and then we expand the integrand in powers of $k$ (up to
$O(k^4)$).
\parindent0.em

\begin{eqnarray}
\label{DF4}
\hspace{-0.5cm}\int\limits_\Lambda d^4p_E\
h\left( {(p_E+k_E)^2\over m^2}\right)\ &=&
\ \int\limits_\Lambda d^4p\ \left\{ \ h(s)
\ +\ 2\ {pk\over m^2}\ h^\prime (s)
\ +\ {k^2\over m^2}\ h^\prime (s)\ +\right.\nonumber\\
\vspace{0.3cm}\nonumber\\
&&+\ 2\ {(pk)^2\over m^4}\ h^{\prime\prime} (s)\ +\
2\ {k^2\ pk\over m^4}\ h^{\prime\prime} (s)\ + \nonumber\\
\vspace{0.3cm}\nonumber\\
&&+\ {4\over 3}\ {(pk)^3\over m^6}\ h^{\prime\prime\prime} (s)\ +\
{1\over 2}\ {(k^2)^2\over m^4}\ h^{\prime\prime} (s)\ +\nonumber\\
\vspace{0.3cm}\nonumber\\
&&\left. +\ 2\ {k^2(pk)^2\over m^6}\ h^{\prime\prime\prime} (s)\ +\
{2\over 3}\ {(pk)^4\over m^8}\ h^{\prime\prime\prime\prime} (s)\ +\
\ldots\ \right\}
\end{eqnarray}

For convenience, we have omitted the subscript $E$ on the
right hand side. Deleting in the integrand terms
antisymmetric with respect to $p \longrightarrow -p$ and applying
following equivalences (valid under the 4D integral)

\begin{eqnarray}
\label{DF5}
(pk)^2\ &{\displaystyle\hat{=}}&\ {1\over 4}\ k^2\ p^2\ \ \ \ ,\\
\label{DF6}
(pk)^4\ &{\displaystyle\hat{=}}&\ {1\over 8}\ (k^2)^2\ (p^2)^2
\end{eqnarray}

we find after some manipulations

\begin{eqnarray}
\label{DF7}
\Gamma_G [e^{-1} k]\ &=&\ {V_4\over 8\pi^2}\ m^4\
\left\{\ \int_0^{\Lambda^2\over m^2} ds\ s\ h(s)\
-\ {1\over 2}\ {k^2\over m^2}\ \left[ \ s^2\ h^\prime (s)\
\right]_0^{\Lambda^2\over m^2}\ \right. +\ \nonumber \\
\vspace{0.3cm}\nonumber\\
&&\hspace{0.7cm}\left. +\ {1\over 12}\ {(k^2)^2\over m^4}\
\left[ \ 3\ s^2\  h^{\prime\prime} (s)\ +\ s^3\
h^{\prime\prime\prime} (s)\ \right]_0^{\Lambda^2\over m^2}\
+\ \ldots\ \right\}
\end{eqnarray}

where $k_\mu$ denotes the constant (Minkowski space) gauge
potential. A comparison of the second term with Eq.\ (\ref{ZA6})
shows that both mass term results although obtained by different methods
agree as expected. Also the first two terms of Eq.\ (\ref{ZA7}) can
be re-identified in Eq.\ (\ref{DF7}).\\

\parindent1.5em

We may now ask ourselves which conditions are to be placed
on $a$ and $b$ in order to make the mass term vanish when
lifting the regularization. From Eq.\ (\ref{ZA6})
(and the second term in Eq.\ (\ref{DF7})) we see that the
requirement of gauge invariance (i.e., vanishing of
any mass term) yields that $h(s)$ should behave for
$s \longrightarrow \infty $ like
\parindent0.em

\begin{equation}
\label{DF8}
h(s)\ \ \stackrel{s \longrightarrow\infty }{\sim }
\ \ const.\ +\ O\left(s^\kappa\right)
\hspace{0.5cm},\hspace{0.5cm}
\kappa\ < \ -1 \ \ .
\end{equation}

Above condition obviously is also appropriate
to make vanish all higher (in powers of $k$) gauge
non-invariant structures in Eq.\ (\ref{DF7}).
By translating information contained in (\ref{DF8})
one finds following conditions sufficient to obey it
\footnote{We disregard here the somewhat weaker condition
\begin{eqnarray*}
\tilde a(s)\ \ \stackrel{s \longrightarrow \infty}{\sim}
\ \ s^{-1/2}\ +\ O\left(s^\kappa\right)
\hspace{0.5cm},\hspace{0.5cm}\kappa\ <\ -{3\over 2}\ \ \ \ ,
\end{eqnarray*}
and all other variants requiring some fine tuning between $\tilde a$
and $\tilde b$.}.

\begin{eqnarray}
\label{DF9}
&&\tilde a(s)\ \ \stackrel{s \longrightarrow \infty}{\sim}
\ \ O\left(s^\kappa\right)
\hspace{0.5cm},\hspace{0.5cm}\kappa\ <\ -1\\
\vspace{0.4cm}\nonumber\\ \label{DF10}
&&\tilde b(s)\ \ \stackrel{s \longrightarrow \infty}{\sim}
\ \  const.\ +\ O\left(s^\kappa\right)
\hspace{0.5cm},\hspace{0.5cm}const. \not= 0\ \ ,\ \ \kappa\ <\ -1
\end{eqnarray}

 From these relations one recognizes that $\tilde a$ and
$\tilde b$ should behave differently for
$s \longrightarrow \infty$, i.e., they cannot be identical.
Above consideration explains (in part) the no-go
result obtained by {\sc Chr\'etien} and {\sc Peierls} \cite{f}
which is caused by the inappropriate
factorization property of the kernel of the fermion action in the case of
$\tilde a = \tilde b$. Such an Ansatz is also in
contradiction to results for the fermion self-energy calculated
in lowest order of standard QED perturbation theory where
$\tilde a$ and $\tilde b$ already differ (see, e.g., \cite{jau}).\\

\parindent1.5em

Now, if conditions (\ref{DF9}), (\ref{DF10}) are fulfilled
the expression for the induced gauge field action (\ref{ZA5})
significantly simplifies. Then, the UV cut-off
can be lifted without any problem ($\Lambda\longrightarrow\infty$),
the coefficients $C_0$ and $C_{1s}$ connected with terms
spoiling gauge invariance are vanishing and the
completely gauge invariant result finally reads
\parindent0.em

\begin{eqnarray}
\label{ZA9}
\hspace{-0.5cm}\Gamma_G [A]\ &=&\ const.\ +\nonumber\\
&&\ \ \ +\ \ C_{1a}\ \ {e^2\over 16\pi^2}
\ \int d^4x \  A^\mu(x)\
[g_{\mu\nu} \Box\ -\ \partial_\mu \partial_\nu]\ A^\nu(x)\ +
\ \ldots\ ,\
\end{eqnarray}

with

\begin{eqnarray}
\label{ZA10}
C_{1a}\ &=&\ {2\over 3}\ \ln\ \left[
{\tilde b (\infty)\over\tilde b(0)} \right]^2\ - \nonumber\\
\vspace{0.3cm}\nonumber\\
&&\ \ \ -\ \int\limits_0^\infty ds\
{1\over s \tilde a^2 + \tilde b^2 }\
\left[\ {s\ \tilde a^2 \over s \tilde a^2 + \tilde b^2}\
\left[\ s\ \tilde a \tilde a^\prime + \tilde b \tilde b^\prime\ \right]
\ +\right.\nonumber\\
\vspace{0.2cm}\nonumber\\
&&\hspace{1.5cm}
+\ {2\over 3}\ s^3\ \tilde a \tilde a^{\prime\prime\prime}\ +\
3\ s^2\ \tilde a \tilde a^{\prime\prime}\ +\
{2\over 3}\ s^2\ \tilde b \tilde b^{\prime\prime\prime}\ +\nonumber\\
&&\hspace{1.5cm}+\ 2\ s\ \tilde a \tilde a^\prime\ +\
3\ s\ \tilde b \tilde b^{\prime\prime}\ -\
s\ (\tilde b^\prime)^2\ +\
3\ \tilde b \tilde b^\prime\ \Bigg] \ \ \ .
\end{eqnarray}

It is worth noting that the coefficient $C_{1a}$ is finite
due to conditions (\ref{DF9}), (\ref{DF10}). Gauge invariance
and UV finiteness are closely related here \footnote{Were it
not for the first term ($\sim (s \tilde a^2 + \tilde b^2)^{-2}$)
in the integral in Eqs. (\ref{ZA8}),
(\ref{ZA10}), also the weaker condition given in the footnote on
p. 7 then replacing (\ref{DF9}) would lead to gauge invariance and
UV finiteness at the same time.}.
All further vacuum
polarization terms are of course finite as in standard QED.
To see this note that all coefficients of these terms have
representations (so far not calculated explicitly yet)
analogous to Eqs.\ (\ref{ZA6})--(\ref{ZA8}).
However, for purely dimensional reasons their ingredients
decay faster for $s \longrightarrow \infty$
than those of the latter. This is the cause that they are finite even in
standard QED. The same argument applies to interaction terms
in the induced gauge field action $\Gamma_G [A]$. This can
easily be inferred from the third term (and all further terms) in
Eq.\ (\ref{DF7}). Consequently, if $\tilde a$, $\tilde b$ obey
conditions (\ref{DF9}), (\ref{DF10}) the induced gauge field
action $\Gamma_G [A]$ is completely finite.
It seems plausible that the present result will generalize to
non-abelian gauge theories as well.\\

\parindent1.5em
Let us finally further comment the no-go result of
{\sc Chr\'etien} and {\sc Peierls} \cite{f}. If one
chooses $\tilde a = \tilde b$ (as done in Ref.\ \cite{f}), one immediately
recognizes from Eq.\ (\ref{ZA8}) that the photon polarization
function is logarithmically (at least) divergent then irrespectively
of the particular choice of $\tilde a$ ($= \tilde b$) made.
If one chooses $\tilde a = \tilde b\ \sim \exp[-s]$,
e.g., the divergency problem becomes
even worse compared with standard QED. In contradistinction to
standard QED the mass term in $\Gamma_G [A]$ then would even be
less divergent ($\sim \Lambda^4$)
than the kinetic term ($\sim \Lambda^6$) and all further terms finite in
the local limit ($\tilde a\ =\ \tilde b\ \equiv\ 1$) would
likely acquire a divergency as well. However, while this definitely rules
out the Ansatz $\tilde a = \tilde b$ it does by far not rule out,
as we have seen above, any finite nonlocal quantum electrodynamics
in general.\\

The author thanks D. Robaschik for reading a draft version
of the Letter and for helpful advice, and C. Eberlein for support
in computerized reference retrieval.\\

\end{document}